\documentstyle[epsf,12pt]{article}
\newcommand{\be}{\begin{equation}}
\newcommand{\ee}{\end{equation}}
\newcommand{\bear}{\begin{eqnarray}}
\newcommand{\eear}{\end{eqnarray}}

\newcommand{\bm}[1]{\mbox{\bf #1}}
\renewcommand{\vec}{\bm}
\renewcommand{\thefootnote}{\fnsymbol{footnote}}
\newcommand{\grgl}{\:\hbox to -0.2pt{\lower2.5pt
           \hbox{$\sim$}\hss}{\raise3pt\hbox{$>$}}\:}
\newcommand{\de}{\hat{d}'_t}
\newcommand{\dm}{\hat{\mu}'_t}
\newcommand{\des}{\hat{d}'_t\vphantom{d}^2}
\newcommand{\dms}{\hat{\mu}'_t\vphantom{d}^2}
\begin{document}
\begin{titlepage}
\noindent
\hfill HD--THEP--95--25 \\
\vspace{3.8cm}
\begin{center}
%{\bf\LARGE TOP PRODUCTION}\\
%\bigskip
%{\bf\LARGE IN HADRON--HADRON COLLISIONS}\\
%\bigskip
%{\bf\LARGE AND}\\
%\bigskip
%{\bf\LARGE ANOMALOUS TOP--GLUON COUPLINGS}\\
{\bf\Large {\LARGE T}OP {\LARGE P}RODUCTION}\\
\bigskip
{\bf\Large IN {\LARGE H}ADRON--{\LARGE H}ADRON {\LARGE C}OLLISIONS}\\
\bigskip
{\bf\Large AND {\LARGE A}NOMALOUS {\LARGE T}OP--{\LARGE G}LUON
{\LARGE C}OUPLINGS}\\
\vspace{1cm}
P.\ Haberl, O.\ Nachtmann and A.\ Wilch\\
\bigskip
Institut  f\"ur Theoretische Physik\\
Universit\"at Heidelberg\\
Philosophenweg 16, D-69120 Heidelberg, FRG\footnote[0]{\hspace*{-7mm}
Submitted to Phys.~Rev.~D}\\
\vspace{3.5cm}
{\bf Abstract:}\\[7mm]
\parbox[t]{\textwidth}{
We discuss the influence of anomalous $t\bar tG$ couplings
on total and differential $t\bar t$ production cross sections
in hadron--hadron collisions. We study in detail the effects of a
chromoelectric and a chromomagnetic dipole moment, $d'_t$ and
$\mu'_t$, of the top quark. In the {$d'_t$--$\mu'_t$} plane, we
find a whole region where the anomalous couplings give a zero
net contribution to the total top production rate. In differential
cross sections, the anomalous moments have to be quite sizable to
give measurable effects. We estimate the values of $d'_t$ and
$\mu'_t$ which are allowed by the present Tevatron experimental
results on top production. A chromoelectric dipole moment of the
top violates CP invariance. We discuss a simple CP--odd
observable which allows for a direct search for CP violation
in top production.}
\end{center}
\end{titlepage}
\renewcommand{\thefootnote}{\arabic{footnote}}
\section{Introduction}
The observation of the top quark has recently been reported
by both experimental groups, CDF and D0, working at the
Tevatron $p \bar p$ collider. The latest CDF value for the
top quark mass is $m_t=176\pm 8\pm 10$ GeV \cite{cdf}, while
D0 gives the value of $m_t=199 {+19\atop -21} \pm 22$ GeV
\cite{d0}. At the Tevatron, top quarks are pair produced in
$p\bar p$ collisions at a c.m.\ energy of $\sqrt{s}=1.8$ TeV.
Based on an integrated luminosity of 67 ${\rm pb}^{-1}$,
the CDF result for the total cross section for this reaction is
found to be \cite{cdf}
\be
\sigma_{\rm exp}(p\bar p\rightarrow t\bar tX)=
6.8 {\textstyle {+3.6\atop -2.4}}\; {\rm pb}\;.
\label{sigex}
\ee
The D0 collaboration obtains from a data sample corresponding to
50 ${\rm pb}^{-1}$ the cross section \cite{d0}
\be
\vspace*{1mm}
\sigma_{\rm exp}(p\bar p\rightarrow t\bar tX)=
6.4 \pm 2.2\; {\rm pb}\;.
\ee
Both central values are higher than the best theoretical prediction
\cite{LSvN}
\be
\vspace*{1mm}
\sigma_{\rm th}(p\bar p\rightarrow t\bar tX)=
4.79 {\textstyle {+0.67\atop -0.41}}\; {\rm pb}
\quad{\rm for}\quad m_t=176\;{\rm GeV}\;,
\label{sigth}
\ee
obtained from a ${\cal O}(\alpha_s^3)$ Standard Model (SM)
calculation including a resummation of the leading soft gluon
corrections to all orders of perturbation theory.
The electroweak corrections are known to be small, of the order
of a few percent \cite{ewcorr}. For higher values of $m_t$, the
theoretical cross section is even lower.
Taking the values in Eqs.\ (\ref{sigex})--(\ref{sigth})
literally we obtain for a possible anomalous contribution
to the cross section
\par
\hfill\parbox{13.4cm}{
\begin{eqnarray*}
\Delta\sigma_{\rm exp}&\le&\sigma^{\rm mean}_{\rm exp}
-\sigma_{\rm th}+\sqrt{(\delta\sigma^{\rm mean}_{\rm exp})^2
+(\delta\sigma_{\rm th})^2}\\[2mm]
&\simeq& 1.8 {\textstyle {+2.9\atop -2.4}}\; {\rm pb}\;.
\end{eqnarray*} }
\hfill\parbox{0.8cm}{\bear \label{expall}\eear }
\par\noindent
Thus the presently available experimental and theoretical
information allows a rather large anomalous contribution.
\par
Future experimental runs will increase the number of
produced $t\bar t$ pairs, allowing the comparison of
differential cross sections with theory [5--7].
%\cite{diffdib1-diffdib3}.
One will be able to investigate also many other observables,
e.g.\ CP--odd ones.
{}From these measurements one expects to obtain detailed
information on the couplings of the top quark.
This might provide a further confirmation of the Standard Model
or open a window to new physics.
\par
Of special interest is the study of CP--odd observables in top
production and other channels of $p\bar p$ collisions [8--14].
%\cite{BMMN-AKR}.
For theoretical investigations of CP violation
in top production and decay in other contexts we refer to
%\cite{DonV-arens2}
[15--29] and references therein.
\par
In this paper we want to investigate possible effects
of anomalous top--gluon couplings on total and differential
cross sections of the reaction $p\bar p\rightarrow t\bar tX$.
To be specific, we assume the existence of chromoelectric
and chromomagnetic top dipole moments, $d'_t$ and $\mu'_t$.
Although there are stringent experimental bounds on
anomalous contributions to
dipole moments of light fermions, it is not unreasonable to
expect large anomalous
moments for the heavy top quark. One way to
generate these couplings is the exchange
of Higgs scalars in one--loop diagrams
in multi--Higgs extensions of the SM.
Since the top--Higgs coupling is proportional to $m_t$,
the effective dipole coupling can be quite sizable.
\par
In principle, the anomalous couplings $d'_t$ and $\mu'_t$ should
be considered as formfactors, depending on the kinematic
variables of the reaction.
However, in $p\bar p$ collisions at $\sqrt{s}=1.8$ TeV
the $t\bar t$ pairs are produced near threshold, where constant
formfactors should be a good approximation.
Another way to put it is to consider in the framework of
effective Lagrangians an expansion in new coupling terms,
ordered by their dimension. The expansion parameter is then
$1/\Lambda$, with $\Lambda$ the scale of new physics.
The dipole moments $d'_t$ and $\mu'_t$ correspond to the
dimension 5 terms (after symmetry breaking \cite{BernN}),
i.e.\ the terms of order $1/\Lambda$, in the effective
Lagrangian.
\par
Some effects of the top dipole moments $d'_t$ and $\mu'_t$
in the reaction
\be
p\bar p\longrightarrow t\bar tX
\label{rctn}
\ee
have been investigated previously.
In \cite{AAS}, the contribution of both electric and magnetic
moments to the matrix element of the parton reactions
underlying (\ref{rctn}), including final quark
polarisation, was calculated, but only to first order in the
anomalous couplings. The quark spin vectors were then used
for the construction of a CP--odd observable. In Ref.~\cite{BM}
the contribution of $d'_t$ to various CP--odd observables
was studied. In Ref.\ \cite{AKR},
total and differential cross sections were computed
up to fourth order in $\mu'_t$, but for vanishing $d'_t$.
\par
In the following we want to extend the above analyses and
investigate
the combined effects of $d'_t$ and $\mu'_t$ simultaneously.
The outline of our calculation is as follows: We will set the
light quark masses to zero and compute the parton processes
$q\bar q\rightarrow t\bar t$ and  $GG\rightarrow t\bar t$ to
leading order (LO) in QCD, i.e.\ at tree level, but including the
effects of $d'_t$ and $\mu'_t$. Convoluting the parton level
results with the parton distribution functions, we evaluate
the cross section for $p\bar p\rightarrow t\bar tX$ which
depends now, of course, on $d'_t$ and $\mu'_t$:
$\sigma(d'_t,\mu'_t)$. We identify the anomalous cross section
(calculated to LO) as
\be
\Delta\sigma:= \sigma(d'_t,\mu'_t)-\sigma(0,0)\;.
\label{delsig}
\ee
We then add $\Delta\sigma$ to the best next to leading order
(NLO) SM calculation available \cite{diffdib2}.
We note here that higher order QCD effects produce, of course,
a chromomagnetic moment form factor. These effects are
included in the NLO SM calculation and do not concern us here.
Our anomalous moment $\mu_t'$ is understood as the
\underline{additional} piece in the chromomagnetic moment
form factor which may be there due to \underline{new} couplings.
Similarly, higher order electroweak corrections in the SM will
produce a chromoelectric dipole form factor, which, however,
is estimated to be unmeasurably small. Thus, a sizable
chromoelectric dipole form factor must come from
physics beyond the SM.
\par
In \cite{diffdib2} it was shown that single top differential
cross sections from  the full NLO calculation in the SM can well
be approximated by a multiplication of the LO Standard Model
result with a constant factor between 1.4 and 1.6.
Such a constant factor drops out in normalized
differential distributions. Thus the effects of new couplings in
differential distributions calculated in LO as described below can
hardly be masked by NLO SM effects. Finally we
discuss the sensitivity of a simple CP--odd observable
from \cite{BM} to the chromoelectric dipole moment $d'_t$.
\section{The model}
We work with the following effective top--gluon interaction
Lagrangian:
\be
{\cal L}_{t\bar tG}=-g_s\,\bar t\gamma^\mu G_\mu t
-i\frac{d_t'}{2}\;\bar t\sigma^{\mu\nu}\gamma_5G_{\mu\nu}t
-\frac{\mu_t'}{2}\;\bar t\sigma^{\mu\nu}G_{\mu\nu}t\;.
\label{LttG}
\ee
Here $g_s$ is the strong coupling constant, $\mu'_t$ and $d_t'$
are the chromomagnetic and chromoelectric dipole moments,
$\sigma^{\mu\nu}=\frac{i}{2}[\gamma^\mu,\gamma^\nu]$,
$G_\mu=G_\mu^aT^a$ with the gluon fields $G_\mu^a$ and the
$SU(3)_C$ generators $T^a=\frac{1}{2}\lambda^a$ ($a$=1\ldots 8),
and $G_{\mu\nu}=G_{\mu\nu}^aT^a$ with the gluon field strength
tensors $G_{\mu\nu}^a=\partial_\mu G_\nu^a-\partial_\nu G_\mu^a
-g_sf_{abc}G_\mu^bG_\nu^c$. Since the anomalous operators
have mass dimension 5, we introduce the dimensionless
dipole moments $\de$, $\dm$ via
\be
d_t'=\frac{g_s}{m_t}\de\;,\qquad \mu_t'=\frac{g_s}{m_t}\dm\;,
\label{norm}
\ee
with the top mass $m_t$. Both anomalous dipole moment couplings
are chirality changing; the magnetic moment term is even under
the combined action of charge and parity transformations CP,
while the electric moment is CP--odd. The signs and factors of
$\frac{1}{2}$ are chosen such as to yield the correct
nonrelativistic limits.
In Fig.\ 1 we show the Feynman vertex factors following from Eq.\
(\ref{LttG}); note in particular that due to gauge invariance
there is also a $t\bar tGG$ coupling.
For the coupling of light quarks $q$ to gluons as well as for
the gluon self coupling we take the SM values.
\par
With this input we calculate the differential cross sections
$\hat{\sigma}_{q\bar q}$ and $\hat{\sigma}_{GG}$ for the
parton level processes
\par
\vspace*{-3mm}
\hfill\parbox{13.4cm}{
\begin{eqnarray*}
q(q_1)\;+\;\bar q(q_2)&\longrightarrow& t(k_1)\;+\;
\bar t(k_2)\;, \\[2mm]
G(q_1)\;+\;G(q_2)&\longrightarrow& t(k_1)\;+\;\bar t(k_2)\;,
\end{eqnarray*} }
\hfill\parbox{0.8cm}{\bear \label{partproc} \eear }
\par\noindent
to lowest order in QCD, as a function of
the usual Mandelstam variables
\be
\hat{s}=(q_1+q_2)^2\;,\quad
\hat{t}=(q_1-k_1)^2\;,\quad
\hat{u}=(q_1-k_2)^2\;.
\ee
For the quark annihilation, there is only the $\hat s$-channel
diagram shown in Fig.\ 2a (the corresponding $\hat t$- and
$\hat u$-channel diagrams are absent, since we set the top
distribution in the proton and antiproton to zero). The result
has the form
\be
\frac{d\hat{\sigma}_{q\bar q}}{d\hat{t}}=
\frac{\pi\alpha_s^2}{\hat{s}^2}
\frac{8}{9}
\left(\frac{1}{2}-v+z+2\dm+(\dms-\des)+
(\dms+\des)\frac{v}{z}\right)\;,
\label{dsqq}
\ee
where we used the abbreviations
\be
z=\frac{m_t^2}{\hat s} \;,\qquad
v=\frac{1}{\hat{s}^2}(\hat t-m_t^2)(\hat u-m_t^2)
\ee
with the kinematical limits $0\le z\le\frac{1}{4}$,
$z\le v\le\frac{1}{4}$.
The variable $v$ can be expressed in terms of the emission angle
$\hat\vartheta$ of the top quark in the parton c.m.\ system as
\be
v=\frac{1}{4}(1-r^2\cos^2\hat\vartheta)\;,\qquad r=\sqrt{1-4z}\;.
\ee
For the gluon fusion process we have to consider the four diagrams
in Fig.\ 2b--e, and we find
\par
\hfill\parbox{13.4cm}{
\begin{eqnarray*}
\frac{d\hat{\sigma}_{GG}}{d\hat{t}}&=&
\frac{\pi\alpha_s^2}{\hat{s}^2}\frac{1}{12}
\bigg[ \left(\frac{4}{v}-9\right)\left(\frac{1}{2}-v
+2z(1-\frac{z}{v})+2\dm(1+\dm)\right)\\[2mm]
+\;(\dms+\!\!\!\!&\des&\!\!\!\!\!)\left(\frac{7}{z}(1+2\dm)+
\frac{1}{2v}(1-5\dm)\right) + (\dms+\des)^2\left(-\frac{1}{z}
+\frac{1}{v}+\frac{4v}{z^2}
\right)\bigg]\,.
\end{eqnarray*} }
\hfill\parbox{0.8cm}{\bear\label{dsgg} \eear }
\par\noindent
In the limit $\de=\dm=0$, we recover the well known SM results
\cite{dmu0}. We also checked against Ref.\ \cite{AKR} for
the case $\de=0$, where we disagree partly\footnote{
Apparently there are some misprints in the formulae of
\cite{AKR}. In Eq.\ (2) (quark annihilation) there
is a factor of $\beta^2$ missing in the last term, as well as
an overall factor of 4. In Eq.\ (6) (gluon fusion) we agree
with the terms $T_1$--$T_4$, but not with the SM
contribution $T_0$ (in the second factor there is a
term $32x^2$ missing). Note also that $\kappa$ defined in
\cite{AKR} is related to our $\dm$ by $\kappa=2\dm$.}.
At Tevatron energies, tops are produced predominantly via the
annihilation of quarks; the gluon fusion process becomes
important for increasing energy as well as for
higher values of the anomalous dipole moments.
\par
According to the parton model, the cross section for the reaction
$p\bar p\rightarrow t\bar tX$ is obtained from a convolution
of the subprocesses Eq.\ (\ref{partproc})
with parton distribution functions,
\par
\hfill\parbox{13.4cm}{
\begin{eqnarray*}
\!\!\!\!\!&d&\!\!\!\!\!\!\sigma\Big(p(p_1)+\bar p(p_2)
\rightarrow t(k_1)+\bar t(k_2)+X(k_X)\Big)=\\[1mm]
&=&\!\!\!\sum_a\!\int_0^1\!\!dx_1\!\!\int_0^1\!\!dx_2\;
N^{p}_a(x_1)N^{\bar p}_{\bar{a}}(x_2)d\hat{\sigma}_{a\bar a}
\Big(a(x_1p_1)\!+\!\bar a(x_2p_2)\rightarrow t(k_1)\!+\!\bar t(k_2)
\Big)\;,
\end{eqnarray*} }
\hfill\parbox{0.8cm}{\bear \eear }
\par\noindent
where the sum runs over all light quark flavors and the gluons,
$a=u$,$\bar u$,$d$,$\bar d$,$c$,$\bar c$,$s$,$\bar s$,
$b$,$\bar b$,$G$. We evaluate the distribution functions
$N(x,s)$ at the hadron c.m.\ energy $\sqrt{s}$, whereas
the energy $\sqrt{\hat s}$ of the
parton subprocess sets the scale for $\alpha_s$.
In particular, the total cross section
can be written as
\be
\sigma(s)=\sum_a\!\int_0^1\!\!dx_1\!\!\int_0^1\!\!dx_2\;
N^{p}_a(x_1)N^{\bar p}_{\bar{a}}(x_2)\;
\Theta(x_1x_2s-4m_t^2)
\hat{\sigma}_{a\bar a}
(\hat s=x_1x_2s)\;.
\ee
The total parton level cross sections, $\hat{\sigma}_{q\bar q}$
and $\hat{\sigma}_{GG}$, can be calculated analytically
and we include the result for completeness in Appendix A.
\par
Finally we give a useful form of the double differential
cross section with respect to rapidity $y$ and transverse
energy $E_T$ of the $t$ jet,
\be
\frac{d^2\sigma}{dy\,dE_T}=2\sqrt{s}\Delta
\sum_a\!\int_0^1\!\!d\tau\;N^{p}_a(x_1(\tau))
N^{\bar p}_{\bar{a}}(x_2(\tau))\;
\frac{d\hat{\sigma}_{a\bar a}}{d\hat{t}}\;,
\ee
with
\be
\Delta=\frac{\sqrt{s}}{E_T}-2{\rm cosh}y\;,\quad
x_1(\tau)=\frac{1}{1+\tau\Delta{\rm e}^{-y}}\;,\quad
x_2(\tau)=\frac{1}{1+(1-\tau)\Delta{\rm e}^y}\;,\quad
\ee
and the kinematical
limits $m_t\le E_T\le \sqrt{s}/(2{\rm cosh}y)$.
In $d\hat{\sigma}_{a\bar a}/d\hat{t}$ one has to perform
the substitutions
$\hat s\rightarrow x_1(\tau)x_2(\tau)s$ and
$z\rightarrow E_T^2/(x_1(\tau)x_2(\tau)s)$.
This completes our presentation of the formulae we need.
\section{Results}
The numerical evaluation was carried out for the Tevatron,
i.e.\ we consider $p\bar p$ collisions at a c.m.\ energy
of $\sqrt{s}=1.8$ TeV. For the top mass we took 175 GeV.
We used various parton distribution functions (PDF) from
the CERN library PDFLIB, but found only weak dependence on the
actual set; the presented results were computed with the
set HO of Gl\"uck, Reya and Vogt \cite{pdf}.
\par
In Fig.\ 3 we show a contour plot of the anomalous contribution
$\Delta\sigma$ defined in Eq.\ (\ref{delsig}) in the
$\de$--$\dm$ plane. This quantity has a minimal value of
\be
\Delta\sigma_{\rm min}=-2.10\qquad{\rm for}
\quad \de =0\,,\;\;\dm =-0.4
\ee
and increases roughly quadratically with $\de$, $|\dm +0.4|$.
We therefore find a whole region where the contributions
of $\de$ and $\dm$ cancel. The dashed lines include
the experimentally allowed region (cf.\ Eq.\ (\ref{expall})).
Along the solid (dotted) line $\Delta\sigma$ takes the value
1.8 pb (0.0 pb).
As explained before, the anomalous contribution has to be
added to the best theoretical SM value given in Eq.\ (\ref{sigth}).
In this way the theoretical value for the total cross section
$\sigma$ could be lowered down to 2.7 pb. More interestingly,
we see that the limits on $\Delta\sigma_{\rm exp}$ in
Eq.\ (\ref{expall}) allow values of the CP violation parameter
$\de$ up to $\de\approx 1.2$ (at the 1 s.d.\ level) if $\dm$ has
an appropriate size. Thus, large effects of CP violation due
to $\de$ are not excluded by the present information on $\sigma$.
\par
Due to the folding with PDF's, differential cross sections
get smoothened, but still reflect the characteristic features
of the parton level distributions. In Figs.\ 4--6 we show
the normalized differential cross section with respect to the
angle $\vartheta$, the emission angle of the $t$ jet in the
laboratory ($p\bar p$ c.m.) frame. We choose $\vartheta=0$
to correspond to $t$-emission in the direction of flight
of the incoming proton.
In all three plots, the solid lines
are the SM result (in LO). In Fig.\ 4 we compare this to the
distributions obtained with chromoelectric moments $\de =$ 0.2, 0.4,
0.6, 0.8, while Fig.\ 5 shows the effect of a chromomagnetic
moment for the values $\dm =$ 0.2, $-0.2$, $-0.4$, $-0.6$.
In Fig.\ 6 we show the angular distributions if both
anomalous couplings are nonzero, $(\de/\dm) =$ (0.8/$-$0.4)
and (0.4/$-$0.8). These values are chosen such that the
anomalous contribution $\Delta\sigma$ to the total rate
would be undetectable (cf.\ Fig.\ 3).
{}From Figs.\ 4--6 we infer that the anomalous couplings would
have to be rather large in order to be visible with a limited
statistics of $t\bar t$ pairs.
\par
In Figs.\ 7--9 we plot the normalized double differential cross
section with respect to rapidity $y$ and transverse momentum
$p_T$ of the $t$ jet, as a function of $p_T$ for different
values of $y$. Fig.\ 7 shows the influence of chromoelectric
moments $\de= 0.2$ and 0.4, Fig.\ 8 of chromomagnetic moments
$\dm=0.2$ and $-0.2$,  compared to the SM result
(solid line). In Fig.\ 9 we show the combined influence
of chromoelectric and chromomagnetic moments, again for
$(\de/\dm) =$ (0.8/$-$0.4) and (0.4/$-$0.8).
Typically the presence of anomalous dipole moments enhances the
production of $t\bar t$ pairs with high transverse momentum.
\par
As general feature we observe that normalized differential
cross sections are of course more sensitive to anomalous
dipole moments than the total rate.
For small dipole moments, however, measurable differences occur
mainly in phase space regions where the contribution to the total
cross section is small, i.e.\ for large $|\cos\vartheta|$ or
large $p_T$. Only if the anomalous couplings take quite sizable
values, one can expect clear signals.
A shift of the maximum of the curves in Fig.\ 9 of $\approx$
50 GeV in $p_T$ when going from the SM to $(\de/\dm) =$
(0.8/$-$0.4) or (0.4/$-$0.8) should clearly be detectable.
By a detailed investigation of the $\cos\vartheta$ and $y$-$p_T$
distributions we found that they are mainly influenced by the
chromomagnetic moment $\dm$. For fixed $\dm$ we found only little
dependence on $\de$ when varying this quantity in the range
allowed by the total cross section measurement (Fig.\ 3).
\par
Finally we discuss the CP--odd observable $\hat{O}_L$ studied in
\cite{BM} which is directly sensitive to $\de$. The observable
is constructed for the production and decay sequence
\par
\hfill\parbox{13.4cm}{
\begin{eqnarray*}
p+\bar p&\longrightarrow&t+\bar t+X\;,\\
t&\longrightarrow&W^++b\longrightarrow \ell^++\nu_\ell+b\;,\\
\bar t&\longrightarrow&W^-+\bar b\longrightarrow
\ell^-+\bar{\nu}_\ell+\bar b\;,
\end{eqnarray*} }
\hfill\parbox{0.8cm}{\bear \label{chain}\eear }
\par\noindent
where $\ell=e,\mu,\tau$. Let $\vec{p}$ ($\vec{q}_+,\vec{q}_-$)
be the momentum of the proton ($\ell^+,\ell^-$) in the
$p\bar p$ c.m.\ system. Then
\be
\hat{O}_L=\frac{1}{m_t^3|\vec{p}|^2}\;\vec{p}\!\cdot\!
(\vec{q}_+\times\vec{q}_-)\;\vec{p}\!\cdot\!
(\vec{q}_+-\vec{q}_-)\;.
\label{OL}
\ee
This is an observable of the tensor type $T_{ij}$ introduced
in \cite{BLMN} and used in search for CP violation in the decay
$Z\rightarrow \tau^+\tau^-$ \cite{CPtau}. In \cite{BM} the
expectation value of $\hat{O}_L$ was calculated for $m_t=130$ GeV
keeping only terms of zeroth and first order in $\de$ and
setting $\dm=0$. The result was
\be
\langle\hat{O}_L\rangle=-0.012 \de\;.
\ee
The number of events (Eq.\ (\ref{chain})) needed to see
a 1 s.d.\ effect can be estimated as
(cf.\ Eq.\ (3.2) of \cite{BM})
\be
N\simeq\frac{50}{|\de|^2}\;.
\ee
If we take these numbers as an indication of the sensitivity
of $\hat{O}_L$ to $\de$ also for the top mass
$m_t=175$ GeV we can conclude that a few thousand events of
the type Eq.\ (\ref{chain}) will be needed to see an effect
of $|\de|=0.1$ at the 1 s.d.\ level by a measurement of
$\langle\hat{O}_L\rangle$. A calculation of
$\langle\hat{O}_L\rangle$ and of the expectation values of other
CP--odd observables for $m_t=175$ GeV and including the effects
of $\de$ and $\dm$ to all orders is in progress.
\section{Conclusions}
In this paper we have investigated the combined effects
of a chromoelectric and chromomagnetic dipole moment
of the top quark on the reaction $p\bar p\rightarrow t\bar tX$.
We have calculated the matrix elements for the parton subprocesses
$q\bar q\rightarrow t\bar t$ and $GG\rightarrow t\bar t$
in leading order QCD.
The numerical evaluation of total and differential cross sections
was done for Tevatron energies.
\par
Our main findings can be summarized as follows:
\begin{itemize}
\item[(1)] In the total cross section, a combination of
chromoelectric and chromomagnetic dipole moments can yield
a positive, negative or zero contribution.
The total rate allows substantial values for the
dipole moments: $\de$, $\dm$ of order 1.
\item[(2)] Differential distributions can discriminate between
chromoelectric and chromomagnetic dipole moments. However,
since the main effect of the anomalous couplings is to change
the absolute rate, normalized differential cross sections are
sensitive only to quite sizable values of the couplings.
\end{itemize}
The most promising way to disentangle the dipole moments
is to exploit their different transformation behaviour
under CP. Since a chromoelectric (chromomagnetic) dipole moment
is odd (even) under CP, one can construct CP--odd observables
which are then sensitive to the chromoelectric moment
$\de$ only. A further advantage is that a CP--odd observable has
already a linear dependence on $\de$, whereas in cross sections
$\de$ can occur only with even powers. The CP--odd observable
$\hat{O}_L$, Eq.\ (\ref{OL}), should be suitable for such an
investigation.
\section*{Acknowledgments}
The authors would like to thank W.~Bernreuther and A.~Brandenburg
for useful discussions. We also thank T.G.~Rizzo for correspondence.

\newpage
\section*{Appendix A}
\renewcommand{\theequation}{A.\arabic{equation}}
\setcounter{equation}{0}
In this Appendix we derive the total parton level
cross sections $\hat\sigma_{a\bar a}$, $a=q,G$. With
\be
\hat{t}=m_t^2-\frac{\hat s}{2}(1-r\cos\hat{\vartheta})\;,
\qquad v=-\left(\frac{\hat t}{\hat s}-z\right)\left(1+
\frac{\hat t}{\hat s}-z\right)
\ee
we have
\be
\hat\sigma_{a\bar a}=\int_{-1}^1d\cos\hat{\vartheta}
\frac{d\hat\sigma_{a\bar a}}{d\cos\hat{\vartheta}}=
\int_{\hat{t}_{\rm min}}^{\hat{t}_{\rm max}}
d\hat t\;\frac{d\hat\sigma_{a\bar a}}{d\hat t}(z,v(\hat t))\;,
\ee
with $\hat{t}_{\rm min/max}=m_t^2-\frac{\hat s}{2}(1\pm r)$.
One can therefore translate the integration to
the simple substitution rules
\be
\hat\sigma_{a\bar a}\,=\,r\hat s\;
\frac{d\hat\sigma_{a\bar a}}{d\hat t}
\left(\begin{array}{lcl}
v^{\phantom{-1}}&\rightarrow&(1+2z)/6\\
v^{-1}&\rightarrow&2L\\
v^{-2}&\rightarrow&2\left(z^{-1}+2L\right)\end{array}
\right)\;,\qquad
L=\frac{1}{r}\ln\left(\frac{1+r}{1-r}\right)\;.
\ee
These rules can be applied directly to the differential cross
sections  $d\hat\sigma_{a\bar a}/d\hat t$ of Eqs.\ (\ref{dsqq}),
(\ref{dsgg}), leading to the result
\be
\hat{\sigma}_{q\bar q}=
\frac{\pi\alpha_s^2}{\hat{s}}\frac{8r}{27}
\left(1+2z+6\dm+2(2\dms-\des)+
\frac{1}{2z}(\dms+\des)\right)\;,
\ee
\par
\hfill\parbox{13.4cm}{
\begin{eqnarray*}
\hat{\sigma}_{GG}\!\!\!&=&\!\!\!
\frac{\pi\alpha_s^2}{\hat{s}}\frac{r}{12}\Bigg[
-7-31z+4L\left(1+4z+z^2\right)\\[2mm]
+2\dm\!\!\!\!\!\!&(&\!\!\!\!\!\!\dm+1)\left(-9+8L\right)+\;
(\dms+\des)\left(\frac{7}{z}(1+2\dm)+L(1-5\dm)\right)\\[2mm]
&&+\;(\dms+\des)^2\left(\frac{1}{3z}\left(1+\frac{2}{z}\right)
+2L\right)\!\Bigg]\;.
\end{eqnarray*} }
\hfill\parbox{0.8cm}{\bear \eear }
\par\noindent
\newpage
\newpage
\section*{Figure Captions}
\begin{description}
\item[Figure 1:] \quad Vertex factors following from
     the top--gluon interaction Lagrangian in Eq.\ (\ref{LttG}).
     All momenta are taken to be ingoing. The necessity of the
     second coupling is a consequence of gauge invariance.
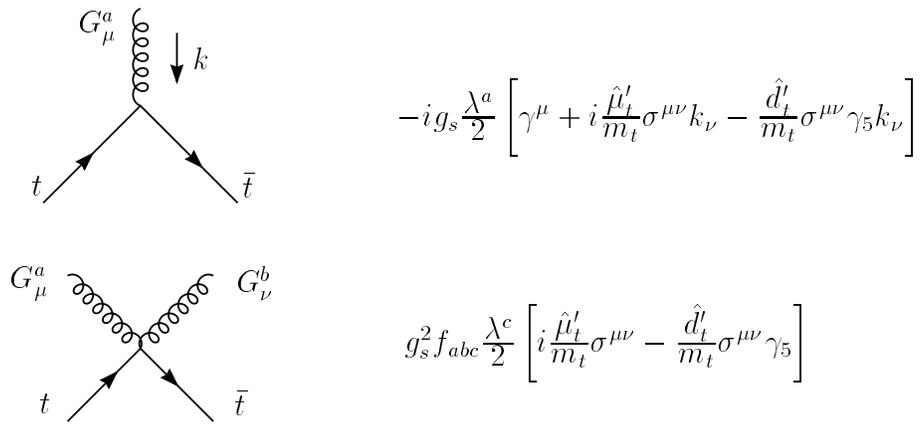
\item[Figure 2:] \quad Feynman diagrams for the quark
     annihilation $q\bar q\rightarrow t\bar t$ (a) and
     the gluon fusion $GG\rightarrow t\bar t$ (b--e) processes.
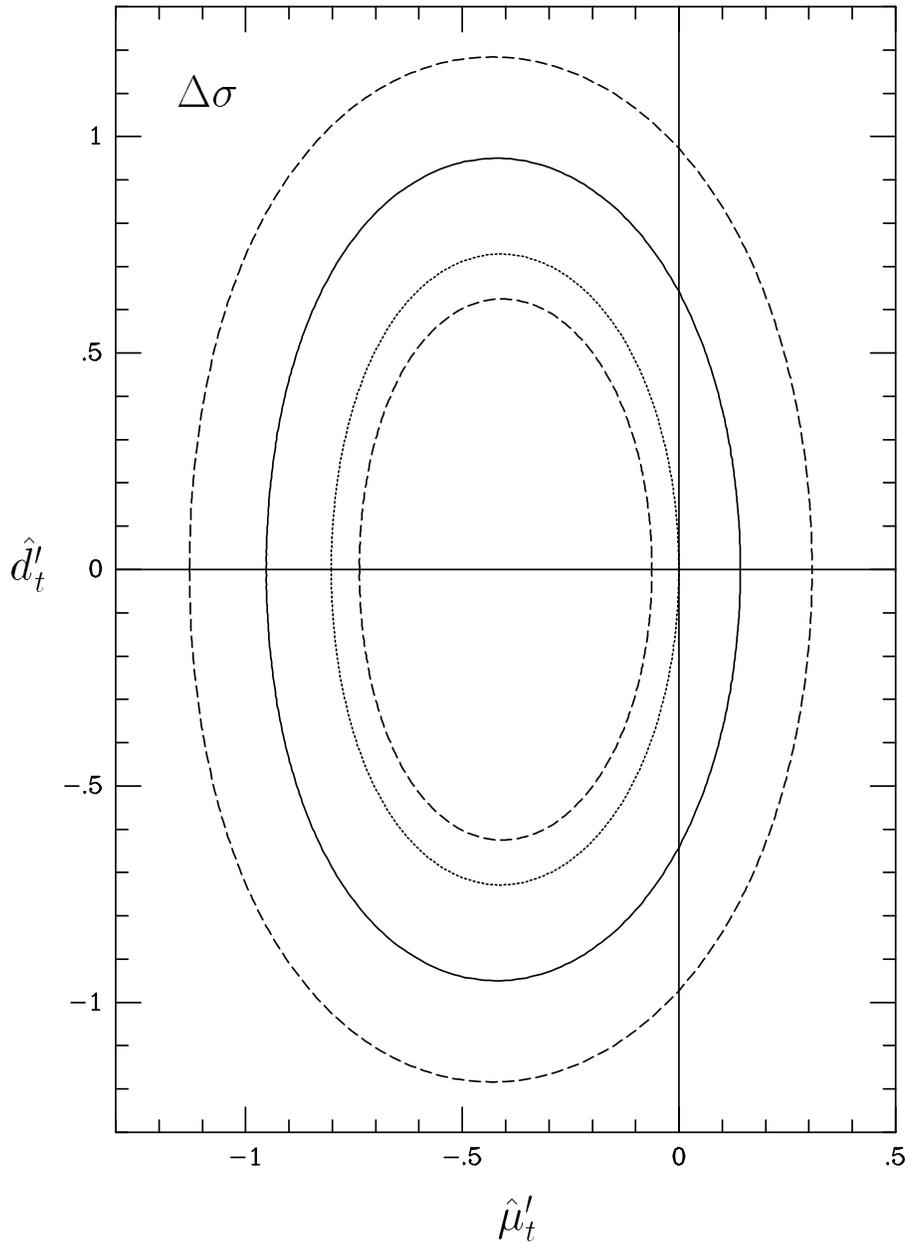
\item[Figure 3:] \quad Contour plot of the anomalous contribution
     $\Delta\sigma$ defined in Eq.\ (\ref{delsig}) as function
     of the chromoelectric and chromomagnetic dipole moments
     $\de$ and $\dm$ (cf.\ Eq.\ (\ref{norm})). The solid line
     corresponds to the mean experimental value $\Delta\sigma= 1.8$
     pb (cf.\ Eq.\ (\ref{expall})). The dashed lines enclose the
     experimentally allowed region (1 s.d.):
     $-0.6$ pb $\le\Delta\sigma\le 4.7$ pb. The dotted line
     corresponds to the SM result $\Delta\sigma= 0$.
\item[Figure 4:] \quad Normalized differential cross section
     $(1/\sigma)(d\sigma/d\cos\vartheta)$ for
     $p\bar p\rightarrow t\bar tX$, where $\vartheta$ is the
     angle of the $t$ jet in the $p\bar p$ c.m.\ frame. The solid
     line represents the LO SM result. The long-dashed
     (short-dashed, dot-dashed, dotted) line shows the effect of
     a chromoelectric dipole moment $\de=$ 0.2 (0.4, 0.6, 0.8)
     with $\dm=0$.
\item[Figure 5:] \quad Same as Fig.\ 4, now for different values
     of the chromomagnetic dipole moment. The long-dashed
     (short-dashed, dot-dashed, dotted) line corresponds to
     $\dm=$ 0.2 ($-$0.2, $-$0.4, $-$0.6) with $\de=0$.
\item[Figure 6:] \quad Same as Fig.\ 4, with nonzero values for
     both anomalous dipole moments. The dashed line shows the
     effect of $(\de/\dm) =$ (0.8/$-$0.4), the dotted line is
     obtained from $(\de/\dm) =$ (0.4/$-$0.8).
\item[Figure 7:] \quad Normalized double differential cross section
     $(1/\sigma)(d^2\sigma/dydp_T^2)$ for the reaction
     $p\bar p\rightarrow t\bar tX$ plotted versus $p_T$ for
     different values of $y$. $p_T$ and $y$ are transverse
     momentum and rapidity of the $t$ jet. Shown is the SM result
     (solid line) and the distributions obtained with an
     anomalous chromoelectric moment $\de=0.2$ (dashed)
     and $\de=0.4$ (dotted) for $\dm=0$.
\item[Figure 8:] \quad Same as Fig.\ 7, now for chromomagnetic
     dipole moments $\dm=0.2$ (dashed) and $\dm=-0.2$ (dotted)
     for $\de=0$.
\item[Figure 9:] \quad Same as Fig.\ 7, now with both $\de$
     and $\dm$ nonvanishing. The dashed line corresponds to
     $(\de/\dm) =$ (0.8/$-$0.4), the dotted line to  (0.4/$-$0.8).
\end{description}
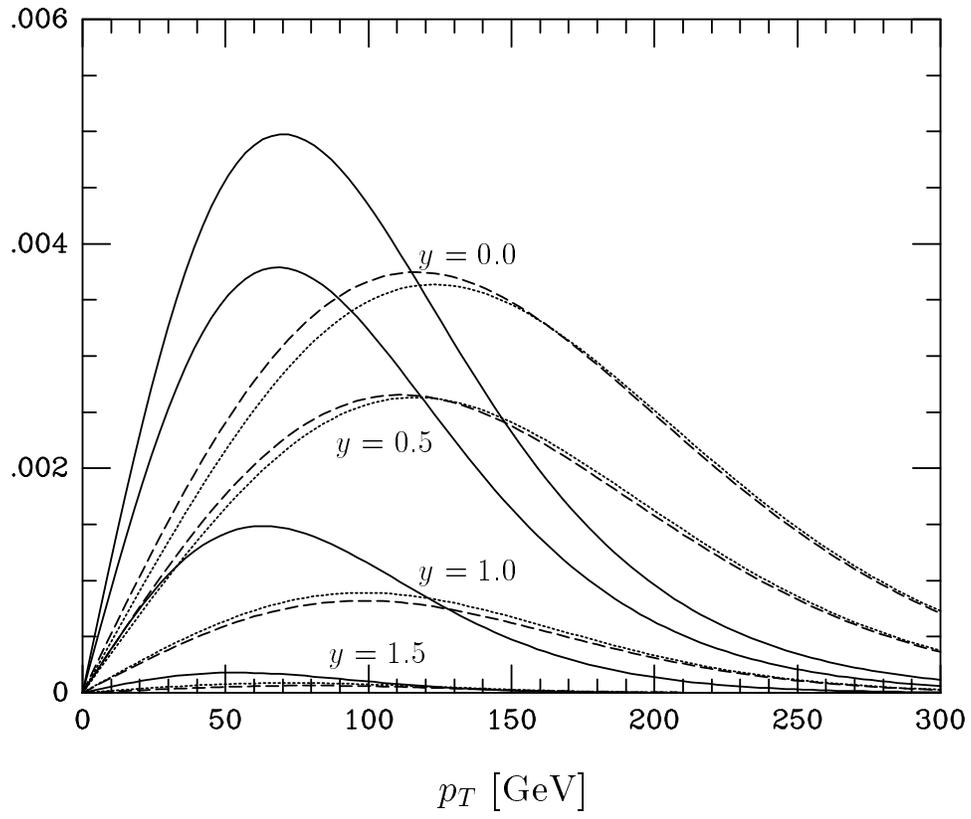
\end{document}